\documentclass[final,5p,times]{elsarticle}
\usepackage{lineno,hyperref}
\modulolinenumbers[5]

\journal{Journal of \LaTeX\ Templates}
\usepackage{multicol}

\usepackage[utf8]{inputenc}
\usepackage{url}
\usepackage{float}
\usepackage{graphicx}
\usepackage{caption}
\usepackage{subcaption}
\graphicspath{./}
\usepackage{color}
\usepackage{dirtytalk}

\begin{document}

\begin{frontmatter}

\title{YSO implantation detector for beta-delayed neutron spectroscopy}

\author[utk]{M.~Singh}

\author[utk]{R.~Yokoyama}

\author[utk,ornl]{R.~Grzywacz}

\author[utk]{A. Keeler}

\author[utk]{T.~T.~King}

\author[ific]{J.~Agramunt}

\author[utk,ornl]{N.~T.~Brewer}

\author[riken]{S.~Go}

\author[riken,hku]{J.~Liu}

\author[riken]{S.~Nishimura}

\author[proteus]{P.~Parkhurst}

\author[vnu,riken]{V.~H.~Phong}

\author[ttech]{M.M. Rajabali}

\author[ornl,utk]{B.C. Rasco}

\author[ornl]{K.P. Rykaczewski}

\author[ornl]{D.W. Stracener}

\author[ific]{A. Tolosa-Delgado}

\author[agile]{K. Vaigneur}

\author[heavyion]{M. Wolińska-Cichocka}

\address[utk]{Department of Physics and Astronomy, University of Tennessee, Knoxville, TN 37996, USA}

\address[ornl]{Physics Division, Oak Ridge National Laboratory, Oak Ridge, TN 37830, USA}

\address[ific]{Instituto de Fisica Corpuscular (CSIC-Universitat de Valencia), E-46071 Valencia, Spain}

\address[riken]{RIKEN, Nishina Center, 2-1 Hirosawa, Wako, Saitama 351-0198, Japan}

\address[hku]{Department of Physics, the University of Hong Kong, Pokfulam Road, Hong Kong}

\address[proteus]{Proteus, Inc., Chagrin Falls, OH 44022, USA}

\address[vnu]{Faculty of Physics, VNU University of Science, 334 Nguyen Trai, Thanh Xuan, Hanoi, Vietnam}

\address[agile]{Agile Technologies, Knoxville, TN 37932}

\address[ttech]{Department of Physics, Tennessee Technological University, Cookeville, TN 38505, USA}


\address[heavyion]{Heavy Ion Laboratory, University of Warsaw, Warsaw PL-02-093, Poland}


\begin{abstract}
A segmented-scintillator-based implantation detector was developed to study the energy distribution of $\beta$-delayed neutrons emitted from exotic isotopes. The detector comprises a 34 $\times$ 34 YSO scintillator coupled to an 8 $\times$ 8 Position-Sensitive Photo-Multiplier Tube (PSPMT) via a tapered light guide. The detector was used at RIBF, RIKEN for time-of-flight-based neutron spectroscopy measurement in the $^{78}$Ni region. The detector provides the position and timing resolution necessary for ion-beta correlations and ToF measurements. The detector provides a high $\sim$ 80 $\%$ beta-detection efficiency and a sub-nanosecond timing resolution. This contribution discusses the details of the design, operation, implementation, and analysis developed to obtain neutron time-of-flight spectrum and the analysis methods in the context of neutron-rich nuclei in the $^{78}$Ni region.
\end{abstract}


\end{frontmatter}

\section{Introduction}


Nuclei with a large neutron-to-proton ratio are known to be produced in stellar environments containing high neutron flux. The beta-decay study of these nuclei is crucial to model the r-process \cite{BURBRIDGE1957} nucleosynthesis. The data on the beta decay of these nuclei is difficult to obtain. However, recent advancements in the radioactive ion-beam facilities have facilitated the production of these neutron-rich nuclei with sufficient statistics to make credible measurements. One of the production methods at the facilities is fragmentation or in-flight fission of an accelerated heavy-ion beam. The method produces a \say{cocktail beam,} often containing several isotopic species. The beam's highly energetic (E$\sim$10 GeV) ions can be stopped in a detector where they undergo beta decay. The daughter nuclei thus produced may result in neutron emission, a process is termed as $\beta$-delayed neutron emission \cite{robert1939}. The initial purpose of the implantation detector is to locate the interaction position in the stopping material of the ions and electrons to establish ion-beta correlations. The precision timing capabilities of the detector are used for ion and electron time measurements for correlations and half-life measurements. The timing properties of the detector can be employed to measure the energy of $\beta$-delayed neutrons using time of flight (ToF) measurements. The energy information of the delayed neutrons is critical for obtaining $\beta$-decay strength distribution ($S_{\beta}$) to neutron-unbound states in the daughter nucleus.

 Currently, several implantation detectors are being used to study radioactive decays at fragmentation facilities, and are typically silicon-based detectors such as AIDA \cite{Griffin2014} and WASA3BI \cite{wasabi}, and were employed at the RIBF, Nishina Center, Japan with the BRIKEN \cite{Tolosa2019} neutron counter to count beta-delayed neutrons. These detectors are excellent for the purpose of ion-beta correlations but are limited in terms of their timing performance. A fast-timing response from the start detector is critical for accurate ToF measurements. 
 
 To achieve a sufficient timing performance for precise time-of-flight (ToF) measurements, the idea of a fast scintillator coupled to a fast-timing Position-Sensitive Photo-Multiplier Tube (PSPMT) was exploited, and a Yttrium Orthosilicate (YSO) based implantation detector was developed\cite{YOKOYAMA201993}. A custom-made segmented YSO scintillator was provided by Proteus Inc. \cite{PROTEUS} for building the detector. The detector was used for the first time at RIKEN Nishina center alongside WASA3BI as an implantation detector to count $\beta$-delayed neutron branching ratios for isotopes in the $\textsuperscript{78}$Ni region in tandem with the BRIKEN neutron counter. The detector consisted of a 48 $\times$ 48 segmented YSO scintillator coupled to an 8 $\times$ 8 segmented PSPMT. The proof-of-principle experiment was successful and demonstrated the YSO detector's capability to measure correlation. It has several advantages over the conventional silicon-based detectors. It provides greater stopping power due to a higher atomic number (Z=39) than Si, leading to a higher beta-detection efficiency. It's much simpler to operate due to the reduced number of electronic channels. 
 
 \section{Detector design and properties}
 
 The YSO-based implantation detector used for ToF measurements spectroscopy is a variant of the YSO detector that was a part of the BRIKEN detector\cite{YOKOYAMA201993}. The new detector version has a larger area for measurements with a larger beam profile.
 
 \begin{figure}[h]
\centering
\includegraphics[width=\columnwidth]{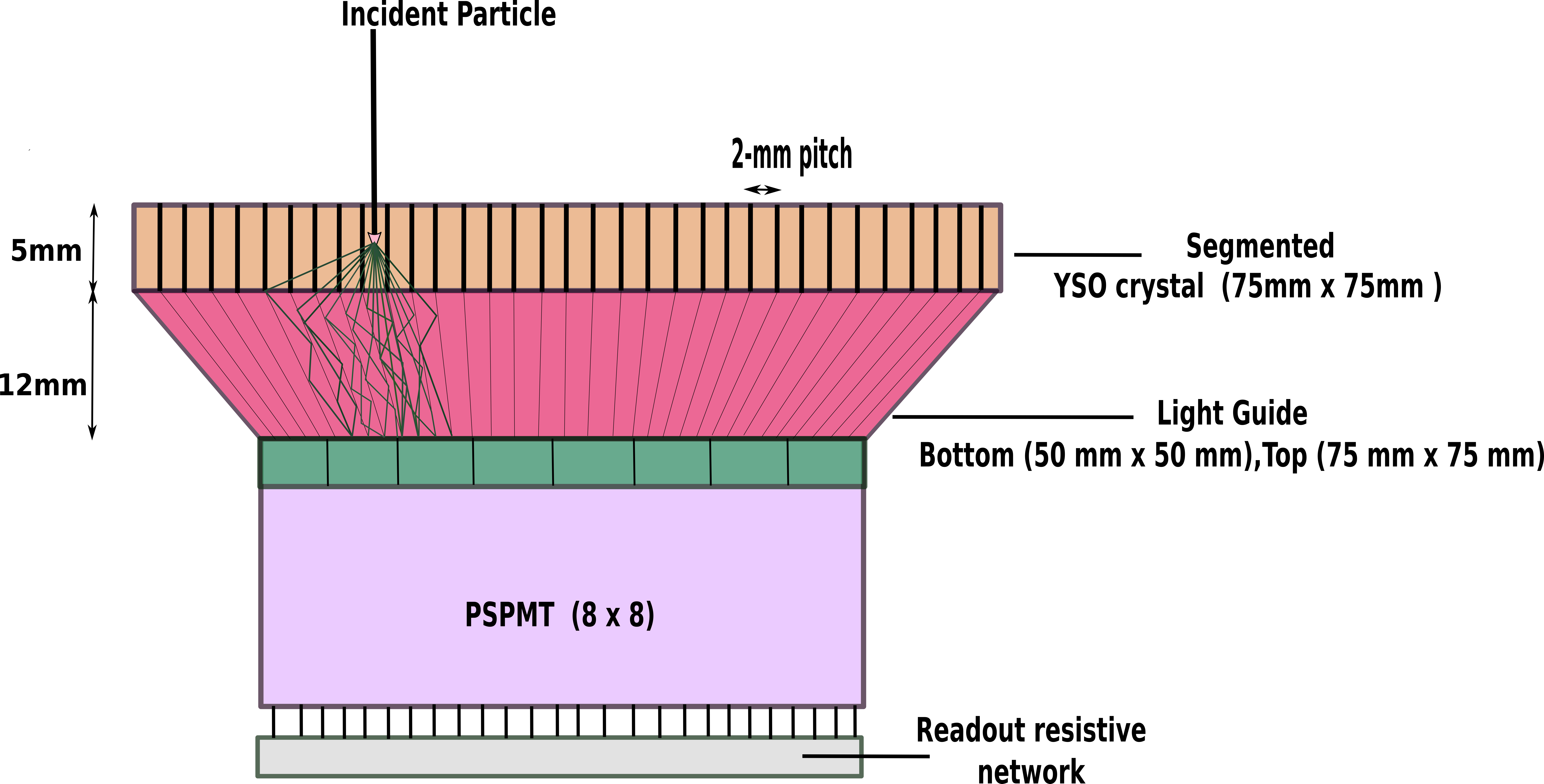}
\caption[]{A graphical layout of the YSO implant detector composed of a segmented YSO scintillator coupled to a PSMPT via an acrylic light guide.}
\label{yso_layout}
\end{figure}

The detector consists of a 75 mm $\times$ 75 mm segmented scintillator with 2-mm pitch coupled to a Hamamatsu H12700B-03 (48.5 mm  $\times$ 48.5 mm effective area) \cite{H12700} via a tapered and pixelated acrylic light guide. The light guide maps the area of the scintillator onto the face of the PSPMT and guides the scintillation light produced in the YSO crystal on the PSPMT entrance window. The sides of the scintillator and the light guide are covered with an Enhanced Specular Reflector (ESR) to contain scintillation light within a pixel. The 64 anode channels of the PSPMT are read out by a resistive network array using the concept called Anger logic \cite{Anger1958}. In this implementation, Vertilon SIB064-1730 (a custom version of the SIB064B-1018) \cite{Vertilon} was used as a read-out board. The Vertilon board divides the voltage signal generated in the photo-multiplier. The weight of the fraction for all the signals is given by the resistance encountered, which depends on the location of the interaction in the crystal. The board has five readouts, four anodes (parts of the voltage signal divided by the resistor network), and one dynode (total voltage induced by the charged particle). The anode signals are used to construct the position distribution of the impinging ions and the emitted electrons using the concept of Anger Logic \cite{Anger1958}. The dynode signal output measures energy loss by ions and electrons in the YSO. The detector provides $\beta$-trigger, the start time of ToF of the delayed neutrons emitted by the neutron-rich precursors. 

Light quenching in the YSO scintillator enables the detector to detect both the ions and the electrons simultaneously. The light quenching in the YSO allows GeV equivalent ion signals to generate signals on the MeV scale. The scintillation light produced by high-energy ions is subjected to quenching as per Birks' relation \cite{birks1964}
\begin{multicols}{2}
\begin{equation}
\frac{dL}{dx}= S\frac{\frac{dE}{dx}}{1+kB\frac{dE}{dr}}
\end{equation}

\begin{equation}
Q_{i}(E)=\frac{L_{i}(E)}{L_{e}(E)}
\end{equation}
\end{multicols}

where B$\frac{dE}{dx}$ denotes the density of excitation centers along the track, k is the quenching factor, kB as a product is known as Birks' Factor. The quenching factor (QF) for ions is the ratio of their light yields relative to that of electrons at the same energy. Generally, QF is a function of the energy and mass of the primary recoiling nucleus,  the chemical composition of the scintillator (including dopants), and its temperature.


Several tests were performed using radioactive sources to check the images for position resolution and the timing performance of the detector at various biasing voltages. Timing performance was checked using voltage signals with different pulse heights representing various energies expected from high-energy electrons in the experiment. 

\begin{figure}[h]
\centering
\includegraphics[width=\columnwidth]{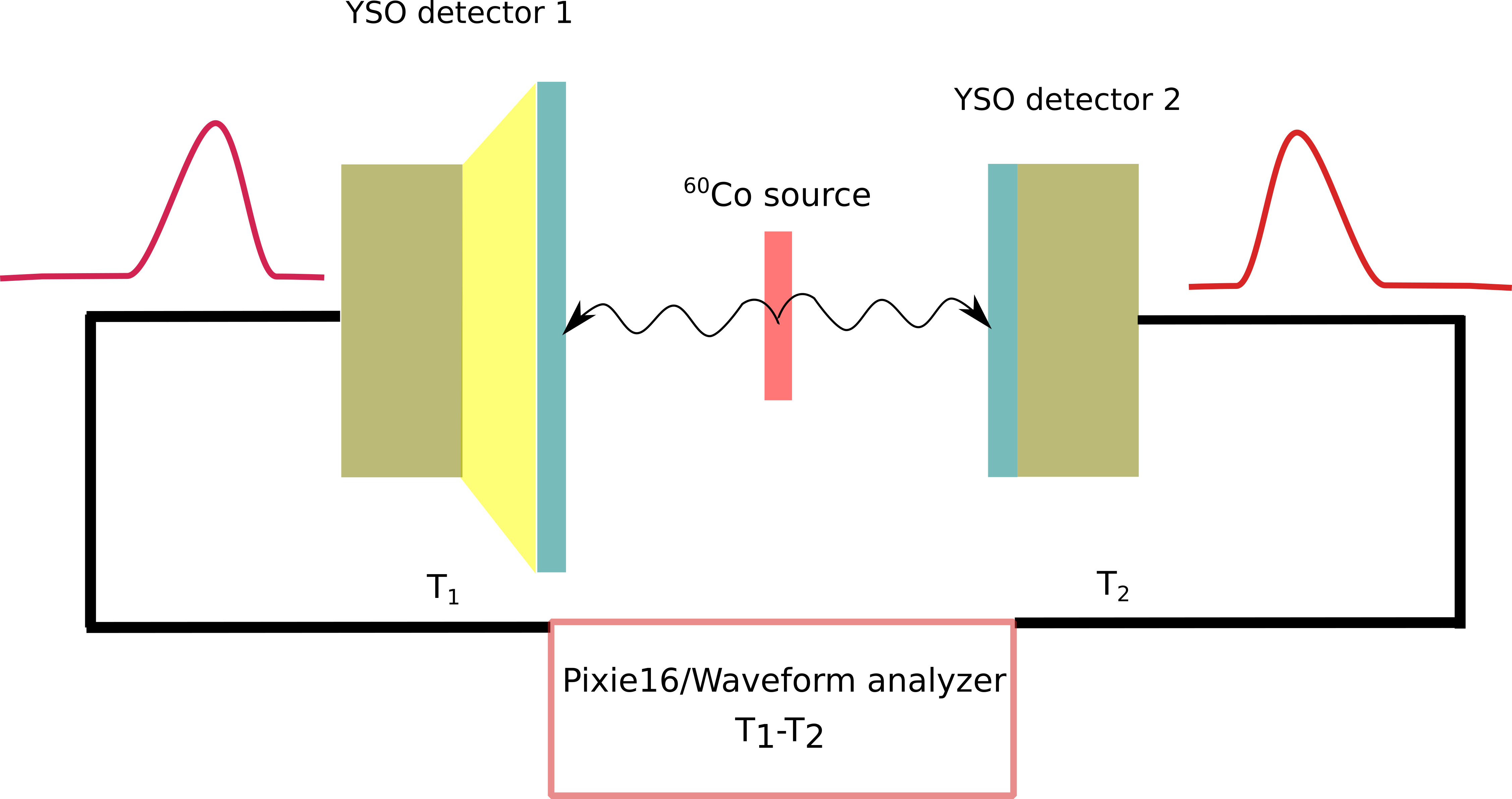}
\caption[timeyso]{Design of the setup to measure timing resolution of the YSO detector.}
\end{figure}

A coincidence time measurement method was used to measure the timing resolution of the detector. The measurement setup involved two similar YSO detectors placed face-to-face with a \textsuperscript{60}Co source in the middle. Beta decay of \textsuperscript{60}Co to \textsuperscript{60}Ni leads to the emission of two coincident gamma lines with energies 1.17 and 1.33 MeV. The timing of the coincident $\gamma$-rays emitted by the source was measured using a pulse shape algorithm in both detectors \cite{PAULAUSKAS201422}. The difference in the timing of the coincident events in both detectors provides a distribution representing the combined timing resolution of the two detectors. The distribution can be approximated to be Gaussian. 

\begin{figure}[h]
\centering
\includegraphics[width=\columnwidth]{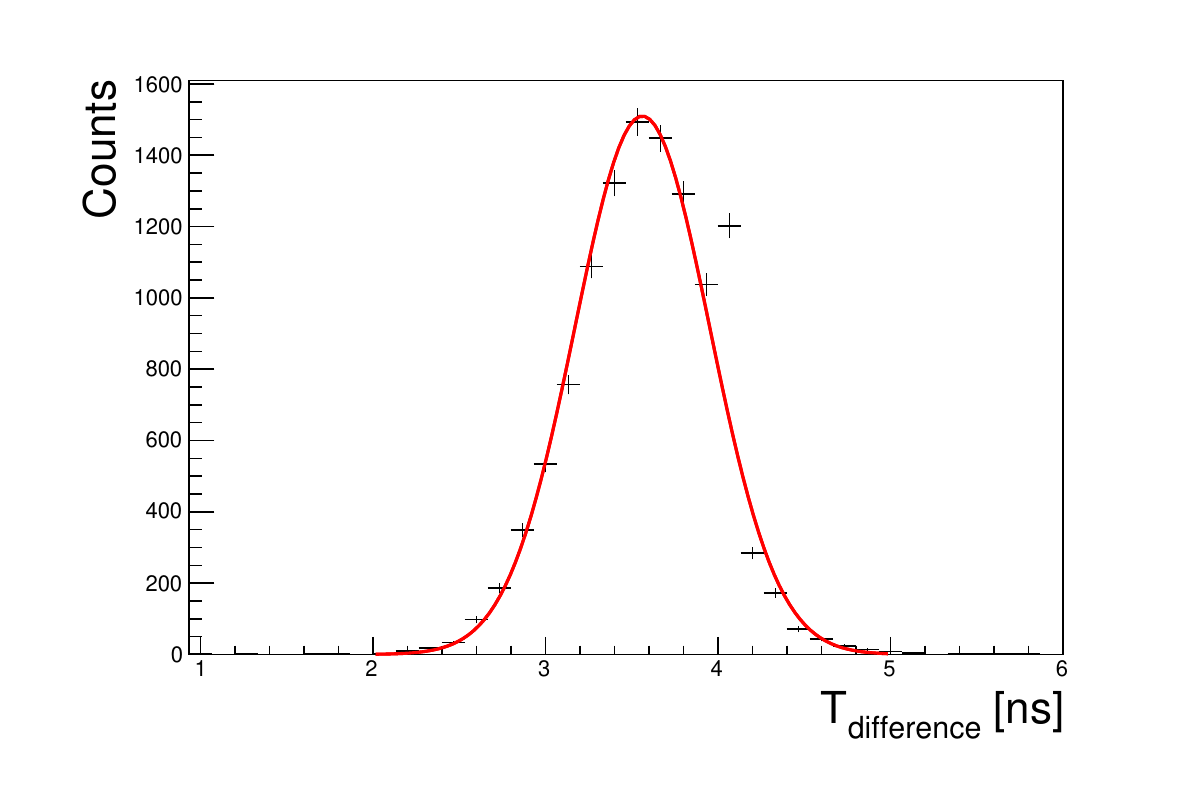}
\caption[timeyso]{Distribution of time difference between the detection of coincident gamma-ray events in the two YSO implant detectors, and fit of the distribution using a Gaussian function. A Full-width Half-Maximum (FWHM) of 922$\pm$7 \emph{ps} was calculated. }
\end{figure}

The Full-Width Half-Maximum (FWHM) ($T_{FWHM}$) of the  distribution  gives  an  estimate of  the  timing resolution  of  both  the  detectors combined as follows: 

\begin{equation}
T_{FWHM} = \sqrt{{T_{res1}}^{2}+{T_{res2}}^{2}},
\end{equation}
where ${T_{res1,2}}$ denotes the timing resolution of the individual detectors. The timing resolution of a single detector can be deduced assuming a replication ($T_{res1} = T_{res2}=T_{res}$) of the timing performance of both detectors. Hence, the resolution the detector ($T_{res}$) can be deduced by measuring $T_{FWHM}$ as follows:

\begin{equation}
T_{res} = \frac{T_{FWHM}}{\sqrt{2}}.
\end{equation}

Scenarios with different biasing voltages were also explored. The detector, on average, has a timing resolution better than $\sim$ 650 ps for signals representing energy greater than or equal to 1 MeV. The timing resolution of the detector encapsulates contributions from transit-time uncertainty in the PSPMT, timing response of the scintillators, and digital electronics.

\section{Experiment}

The experiment to study $\beta$-delayed neutron emissions around \textsuperscript{78}Ni was performed at Radioactive Ion Beam Factor (RIBF) at Riken Nishina Center, Japan. The neutron-rich nuclei were produced by in-flight fission of a primary \textsuperscript{238}U\textsuperscript{86+} beam with an energy of 345 MeV/nucleon, induced at a 4-mm-thick \textsuperscript{9}Be production target. The typical intensity of the primary beam was $\sim$ 42 pnA during the run. Fission fragments were separated and identified in the BigRIPS \cite{Kubo2003} in-flight separator on an event-by-event basis by their proton numbers (Z) and the mass-to-charge ratio (A/Q). These quantities are obtained by measuring the B$\rho$, time of flight (TOF), and energy loss ($\Delta$E) in BigRIPS.

\begin{figure}[h]
\centering
\includegraphics[width=\columnwidth]{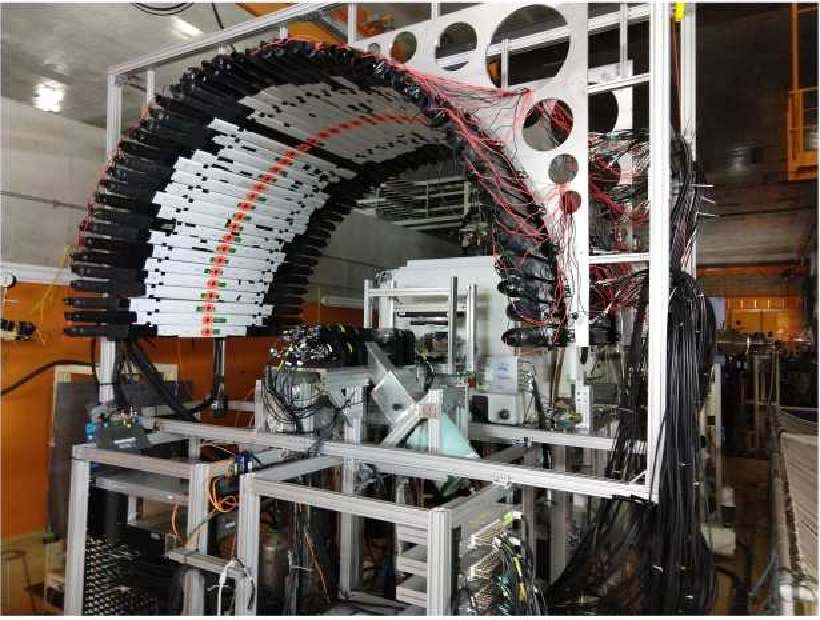}
\caption{Setup of the experiment at the F11 focal plane at RIBF.}
\end{figure}

The setup for the experiment was situated downstream of the F11 focal plane at the facility. The setup for the experiment consisted of 48 medium (3 $\times$ 6 $\times$ 120 cm$^{3}$) VANDLE bars arranged in a 100-cm circular radius with the implantation box containing the YSO detector. The implantation box is a light-tight 3D-printed box containing the YSO detector, a VETO detector, and an auxiliary implant detector. The auxiliary implant detector is an EJ-200 plastic with a volume of ($75 \times 75 \times 5$ $mm^{3}$). The four corners of the plastic are coupled to SiPMs for position reconstruction. The VETO detector consists of a 7-mm-thick EJ-200 plastic with a surface area of ($75 \times 75$ $mm^{2}$). The two diagonal ends of the VETO detector are coupled to SiPMs for readout. The VETO detector is used to gate away the light ions punching through the YSO. In addition, the setup consisted of two High-purity Germanium (HPGe) detectors and ten LaBr\textsubscript{3} (two 2" $\times$ 2" and eight 3\textsuperscript{"} $\times$ 3\textsuperscript{"}) detectors for $\gamma$-rays originating from the decays. All the signals were read using XIA Pixie-16 revF digitizers at 250 MHz and 12-bit digitization\cite{Pixie16}.

\section{Detector setup for the experiment}

For utilizing the YSO detector in the experiment, five signals (4 anodes and 1 dynode) from the detector were split into two sets.

\begin{enumerate}
\item One set of signals was amplified by a factor of 10, called the beta/high-gain branch. The dynode signal from the beta branch was further split into two; one of the split dynode signals was set in coincidence with the VANDLE array, and the other one was fed back to BigRIPS acquisition for time clock synchronization for merging the data streams. 
\item The other set of the five signals from the YSO was labeled as an ion/low-gain branch. These signals were digitized after attenuation. 

\begin{figure}[h]
\centering
\includegraphics[width=7cm, height= 6cm]{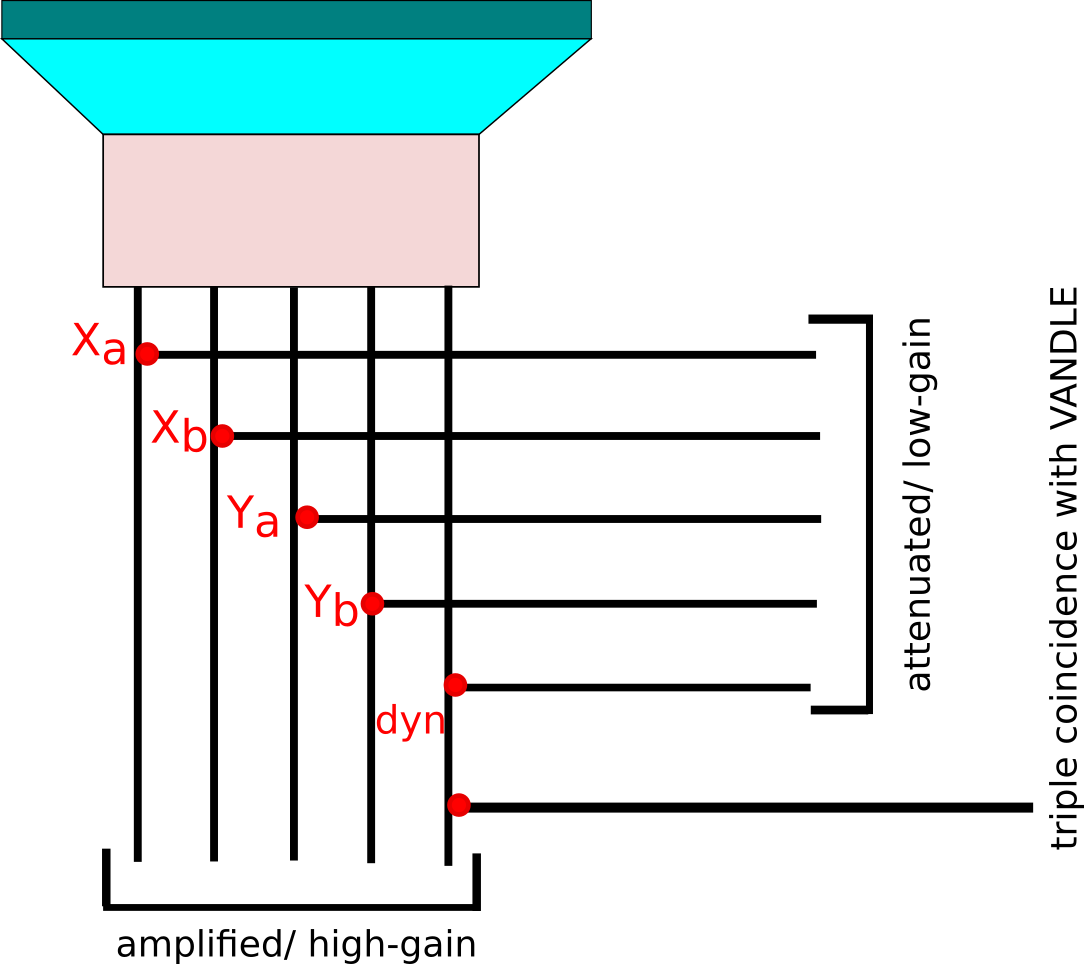}
\caption{Schematic of the trigger scheme adopted for the experiment at RIBF.}
\label{fig:YSO_trigger_riken}
\end{figure}

\end{enumerate}
Digitized traces were saved for all the signals for timing and energy-deposit estimates.

\begin{figure}
\centering
\begin{subfigure}{.2\textwidth}
  \centering
  \includegraphics[width=3cm, height=5cm]{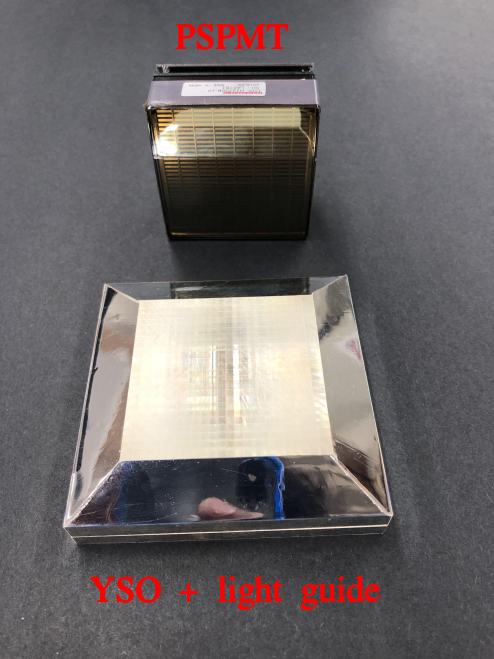}
  \caption{}
  \label{fig:sub1}
\end{subfigure}%
\begin{subfigure}{.2\textwidth}
  \centering
  \includegraphics[width=3cm, height=5cm]{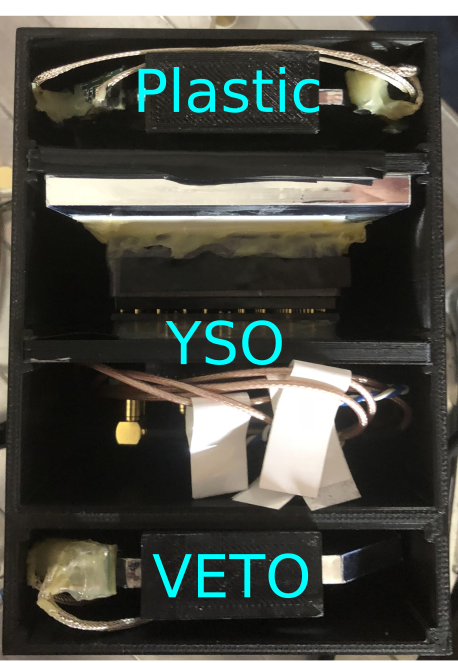}
  \caption{}
  \label{fig:sub2}
\end{subfigure}
\caption{Figure a) shows the pixelated-YSO scintillator coupled with light and the PSPMT, figure b) shows the arrangement of YSO, front implant detector, and VETO in the implantation box.}
\label{fig:test}
\end{figure}

\section{Analysis}
The images for ion and beta distribution in the scintillator were reconstructed using the measured maximum amplitude of the traces of the four dynode signals in high- and low-gain. The images for both ions and betas contain a cross-wire-shaped pattern. The pattern mimics the manufacturing irregularities in the light guide and YSO array due to PSPMT response. The light guide was constructed by connecting four segments of individual light guides for each corner 1/4 fraction of the YSO. The individual segments of the light guide have pixelization matching the YSO. The joints in the light guide affect the light sharing across the pixels of the light guide and hence manifest as a cross-wire-shaped pattern in the final images. The reconstructed images of ions and betas were scaled to match dimensions in the software, and then aligned using the cross wire as a reference.

\begin{figure}[h!]
\centering
\includegraphics[width=\columnwidth]{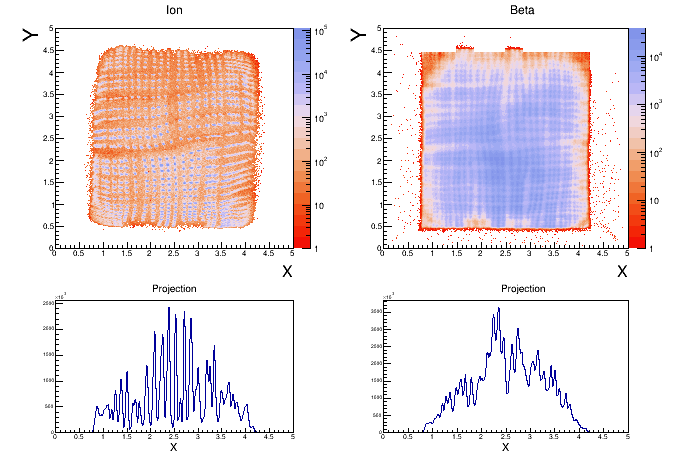}
\caption{Ion- and Beta-position distribution obtained using the Anger logic clearly show the irregularities.}
\label{fig:position_images}
\end{figure}

\begin{figure}[h!]
\centering
\includegraphics[width=\columnwidth]{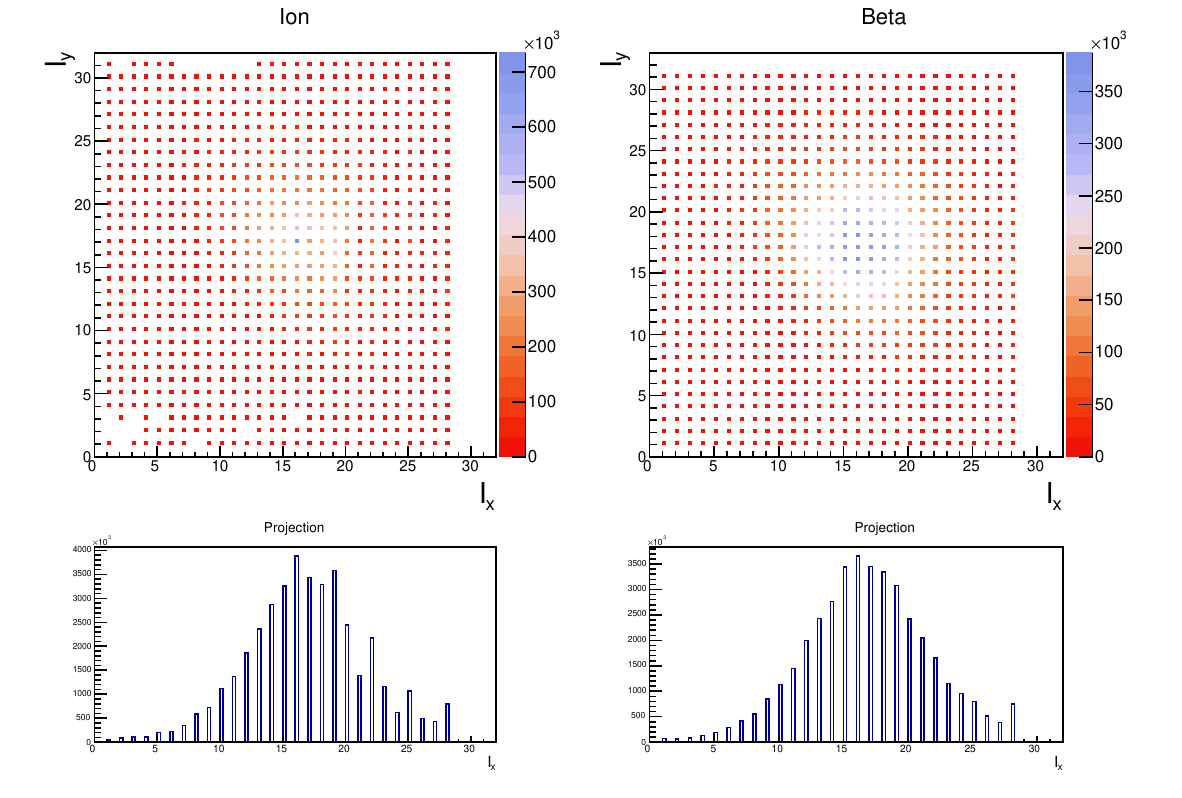}
\caption{Ion- and Beta-position distribution after transforming images to pixel space.}
\label{fig:pixel_images}
\end{figure}

The raw images constructed using the anger logic give a pixel distribution with a varying spatial density, as shown in Fig. \ref{fig:position_images}. The pixel density in the center of the image is sparser, and the density keeps increasing towards the edges of the images in any direction. The non-uniformity doesn't allow for the correlation radius to remain the same. With the raw images, for a given correlation radius, the numbers of pixels correlated would be more around the edges where pixel density is higher as compared to the center where pixel density is lower. Using these images, an accurate description of the variation of $\beta$-detection efficiency with the variation of pixel radius is not achievable.

To establish ion-beta correlations, a pixel map was created containing an array (ID, $X\textsubscript{$\beta$}$, $Y\textsubscript{$\beta$}$, $X\textsubscript{ion}$, $Y\textsubscript{ion}$) of the matching coordinates of the pixel positions in the beta and the ion image in order. Each ion and beta event was allocated a pixel ID in both x and y coordinates using the nearest-neighbor approximation. The pixel IDs were implemented in the algorithm using the following equation:

\begin{equation}
\sqrt{(I_{\beta,i}-I_{ion, k})^2+ (I_{\beta, j}-I_{ion, l})^2} < n.
\end{equation}

Here,  $I_{ion, \beta, i}, I_{ion,\beta,j}$  denote the indices of the pixel associated with a given position of an ion/beta event, and \emph{n} is the number of pixels in the correlation radius.





\begin{figure}[h]
\centering
\includegraphics[width=\columnwidth]{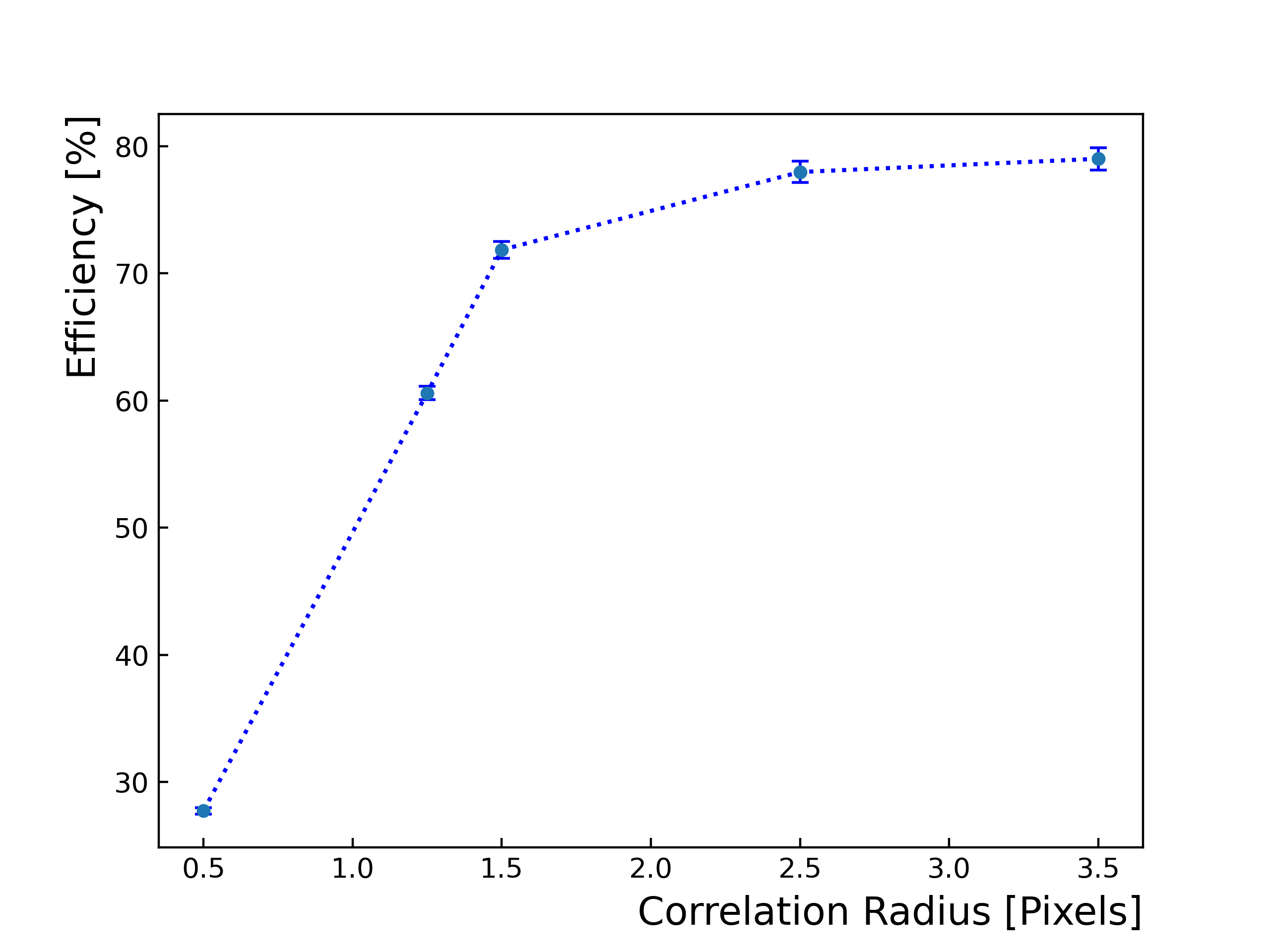}
\caption{$\beta$-detection efficiency of YSO, determined by gating on the decay of $^{79}$Cu}
\label{fig:Cu79_beta_efficieny}
\end{figure}

\begin{figure}[h]
\centering
\includegraphics[width=\columnwidth]{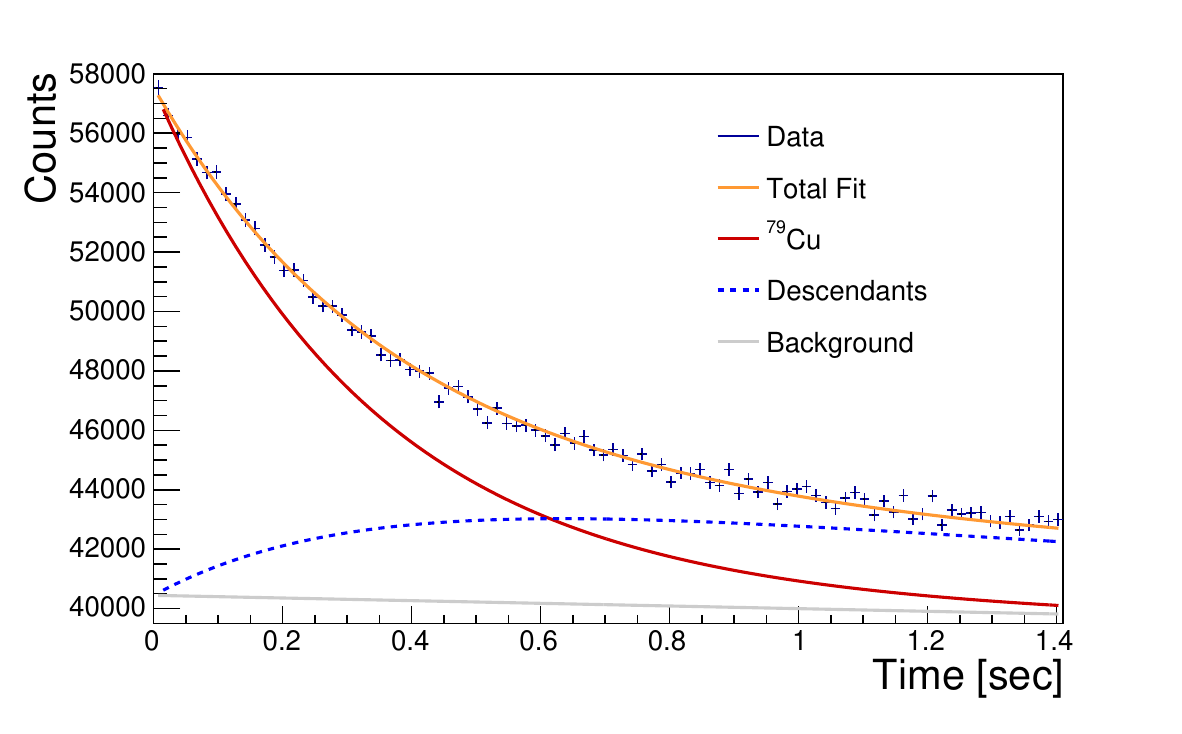}
\caption{$^{79}$Cu decay curve obtained using the ion-beta correlations in pixel space with a correlation radius of 3 pixels. The decay components are fitted using the Bateman Equations.}
\label{fig:Cu79_decay_curve}
\end{figure}

 
 The correlation radius is optimized for maximum beta-detection efficiency specific to the analyzed isotope. Fig. \ref{fig:pixel_images} shows the images obtained after the transformation to find the nearest-neighbor pixel is applied. Using the correlations in the pixel space, the $\beta$-detection efficiency of the YSO detector was calculated. Fig. \ref{fig:Cu79_beta_efficieny} shows the variation of the $\beta$-detection efficiency as a function of pixel radius. A maximum efficiency of $\sim$ 80 $\%$ was obtained in the decay analysis of $^{79}$Cu. Using the correlation criterion, a decay curve, a distribution of the time difference between ion and beta event is achieved by gating on the activity of $^{79}$Cu. The decay curve is shown in Figure \ref{fig:Cu79_decay_curve}. The decay curve is fitted using the Bateman Equations of radioactive decay. The fit gives a half-life of 252.2(9.0) ms, which is in agreement with a previously measured value of 241.0(3.2) ms \cite{Xu2014}.
 
\begin{figure}[h]
\centering
\includegraphics[width=\columnwidth]{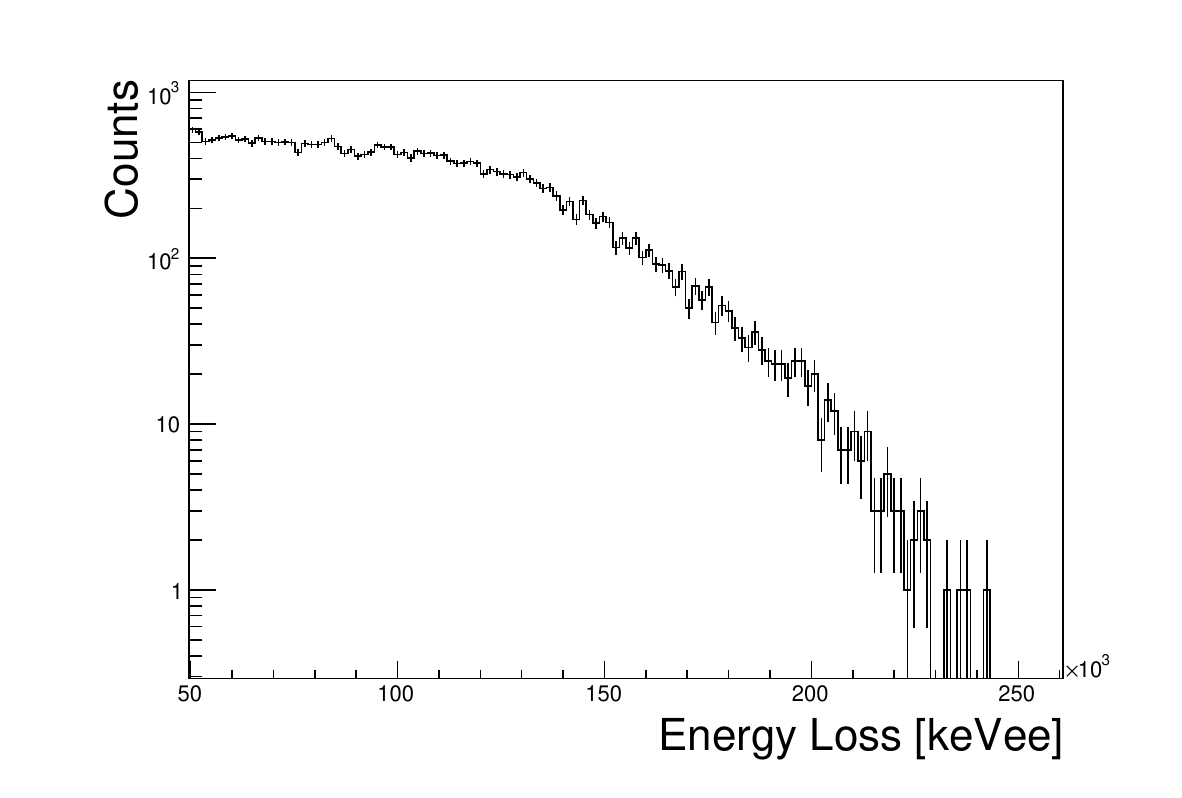}
\caption{Calibrated energy-loss spectrum of $^{82}$Zn in YSO.}
\label{fig:Zn82_figure}
\end{figure}

For measuring the quenching factors, data obtained from the experiment at RIBF as mentioned in \cite{Yokoyama2018} is used for the analysis. The quenching factor was estimated using a semi-empirical approach. To determine the quenching factor for various isotopes, the beta (low-gain) branch of the YSO was calibrated using High-Purity Germanium (HPGe) detectors with a $\textsuperscript{137}$Cs source. The source was placed at the center of the YSO crystal. The $\gamma$-ray with energy 661 keV from the source shares energy loss between the YSO crystal and two HPGe detectors. Using the calibrated HPGe, the high-gain (beta) branch of the YSO detector was calibrated using energy correlations. The dynode spectrum from the ion (low-gain) branch of YSO was then calibrated for the energy of the beta branch by identifying the coincident beta events in the low-gain branch at low QDC. This gives energy of the ions in units of MeVee. Fig.\ref{fig:Zn82_figure} shows the calibrated energy-loss spectrum of $^{82}$Zn in the YSO crystal.

\begin{figure}[h]
\centering
\includegraphics[width=\columnwidth]{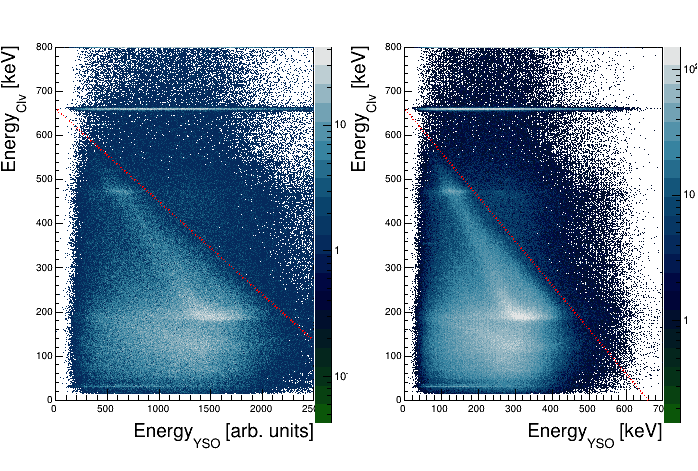}
\caption[]{The correlation between YSO and HPGe detectors before and after calibrating the beta-branch of the YSO detector. The red-dashed lines follow the energy-loss of the 661-keV $\gamma$-rays emitted by the $\textsuperscript{137}$Cs source, shared between the YSO and the HPGe. }
\end{figure}

\begin{figure}[h]
\centering
\includegraphics[width=\columnwidth]{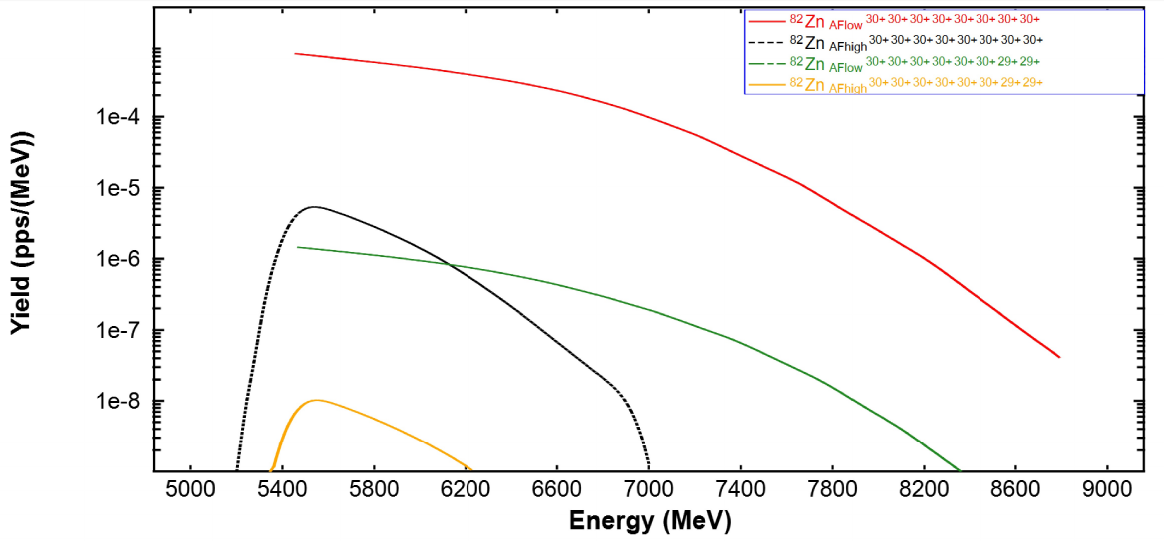}
\caption{Energy-loss distribution of $^{82}$Zn ions in YSO calculated using LISE++. }
\label{fig:Zn82_lise}
\end{figure}


\begin{figure}[h]
\centering
\includegraphics[width=\columnwidth]{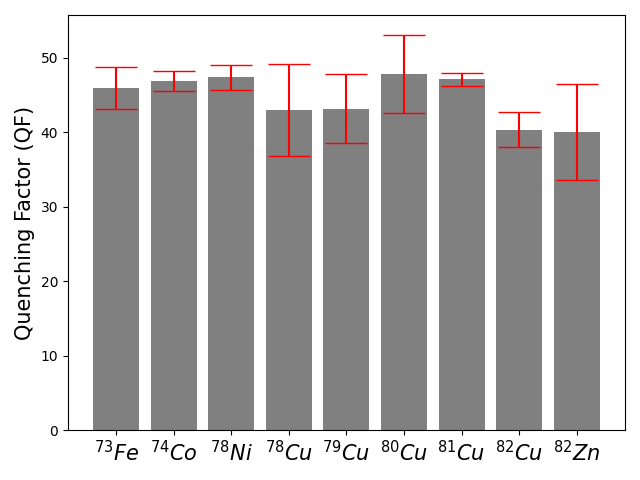}
\caption[]{Light quenching factor determined for various isotopes stopped in the YSO crystal.}
\label{fig:quenching}
\end{figure}

The absolute energy distribution of ions impinging on YSO was simulated using the LISE++ program \cite{TARASOV2016185}. All the detectors and degraders in the beam path were added to properly simulate the energy distribution of the high-energy ions implanted into the YSO. As an example, Fig. \ref{fig:Zn82_lise} shows the energy-loss spectrum calculated using LISE++. The ratio of the end-point energy obtained using the simulated spectrum and the end-point energy in the low-gain branch of the YSO from the data gives an estimate of the quenching of the light produced by the energy loss of the ions in the YSO. Fig \ref{fig:quenching} shows the quenching factors determined for various isotopes stopped in the YSO crystal. The errors in the quenching factor are determined by the error ($\sim$10 $\%$) in the calibration factor for ions and the error in estimating energy from LISE++. The knowledge of light quenching is important for designing experiments with YSO. The estimates of the quenching are important for adjusting the dynamic range for ions with different Z in the software. For Z $\textgreater\textgreater$28, the light quenching is expected to be more severe due to more energy deposit in the scintillator, and the dynamic range of the ions would be shorter compared to the ones in the vicinity of $\textsuperscript{78}$Ni. Whereas, the quenching factor is expected to be smaller for the region of isotopes with Z$\textless\textless$28 due to lesser energy deposit in the scintillator, requiring a broader dynamic range if ions of similar energy distribution as in $\textsuperscript{78}$Ni impinges into YSO. 



%

\section{ToF-based spectroscopy of $\beta$-delayed neutrons} 
Determining $\beta$-delayed neutron energy requires measuring the flight path of neutrons from the point of origin to a point of detection and the time of flight corresponding to the trajectory. The neutron time of flight (ToF) for a given flight path (L) is used to calculate the kinetic energy as shown in the following equations:

\begin{center} 
\begin{equation}
ToF=T_{VANDLE}-T_{YSO},
\end{equation}
\end{center}

\begin{center}
\begin{equation}
 E_{n} = {\frac{1}{2}} {m_{n} \left (\frac{L}{ ToF } \right)}^{2},
\end{equation}
\end{center}
and
\begin{center}
\begin{equation}
\Delta {E} = 2 \times E \times \sqrt{{\left (\frac{\Delta L}{L} \right)}^2+{\left(\frac{\Delta ToF}{ToF}\right)}^2}
\end{equation}
\end{center}

Here, $T_{VANDLE}$ is the stop time determined using the VANDLE array and $T_{YSO}$ is the stop time provided by the YSO implantation detector.

For flight path determination, the position coordinates of the origin of a neutron event in the YSO can be approximated as the position coordinates of the correlated beta event. Further, given the implantation beam spot has a finite dimension and is spread over a diameter of $\sim$ 4 cm, it can lead to miscalculation of the flight path of high energy ($>$ 4 MeV) neutrons, leading to inaccuracies in the kinetic energy evaluation because of the sensitivity to the flight path. Therefore, neutron flight paths were calculated on an event-by-event using the position coordinates of the correlated beta. The position coordinates $(X_{i}, Y_{i}, Z_{i})$ of neutron origin in the YSO are obtained from the position of the correlated beta event using the Anger Logic. The position $(X_{f}, Y_{f}, Z_{f})$ of a neutron event in a VANDLE bar is calculated from the product of the speed of light in the bar and the time difference between the left and right PMT signal. The flight path of a neutron event is calculated using the distance formula for two points in a Cartesian co-ordinates system as follows:

\begin{equation}
L^{'} = \sqrt{(X_{i}-X_{f})^2+(Y_{i}-Y_{f})^2+(Z_{i}-Z_{f})^2}.
\end{equation}

\begin{figure}[h]
\centering
\includegraphics[width=\columnwidth]{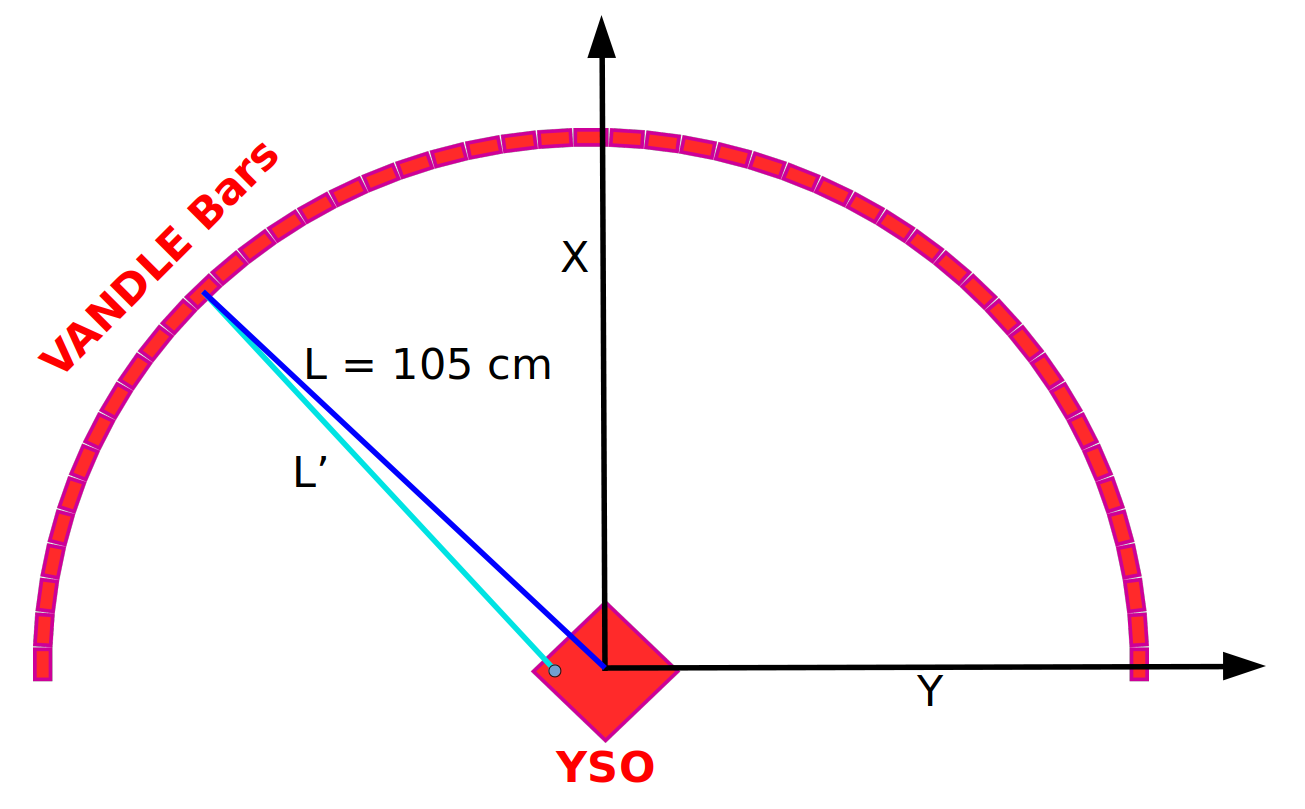}
\caption{Flight path reconstruction scheme for neutron events. All the ToFs for all the flight paths are scaled to ToFs corresponding to a flight path of 105 \textit{cm}. For example, in the graph above, ToF for a neutron event having a flight path of L' is scaled to a ToF corresponding to a flight path L = 105 \textit{cm}. }
\label{fig:Flight_Path}
\end{figure}

The corresponding ToF for a neutron event is calibrated to a flight path of 105 cm, the perpendicular distance between the center of the YSO to the center of a VANDLE bar. A graphical description of the method to correct for the flight path of neutrons based on the YSO dimensions is shown in Figure \ref{fig:Flight_Path}. The $ToF^{'}$ for each flight path ($L^{'}$) is scaled to a ToF corresponding to a flight path of 105 cm as follows:

\begin{equation}
ToF  = ToF^{'}\left ( \frac{105}{L^{'}} \right ).
\end{equation}

\begin{figure}[h]
\centering
\includegraphics[width=\columnwidth]{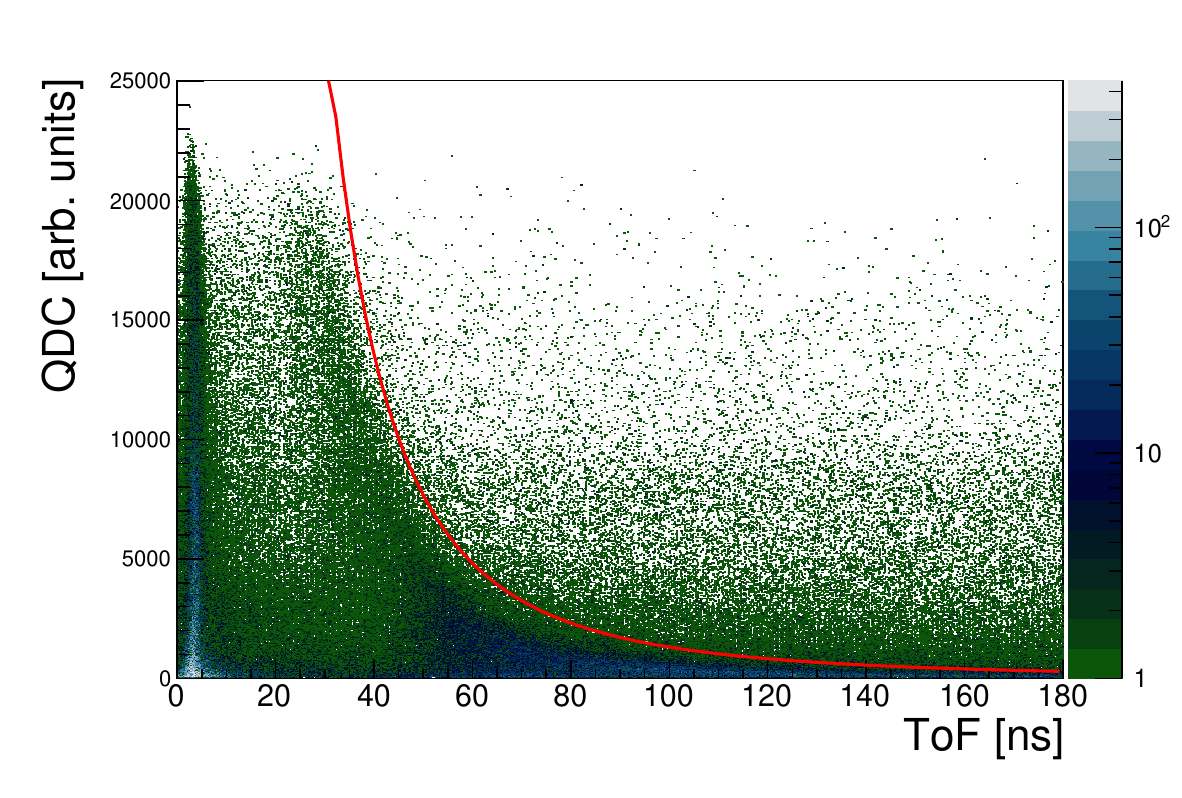}
\caption{QDC vs ToF spectrum of $\beta$-delayed neutrons from $^{83}$Ga obtained using YSO and VANDLE with an ion-beta time correlation window of 300 \textit{ms}.}
\label{fig:Ga83}
\end{figure}

Figure \ref{fig:Ga83} shows the ToF versus Charge-to-Digital Conversion (QDC) spectrum attained for beta-delayed neutrons from the beta-decay of $^{83}$Ga. The spectrum is obtained by restricting to correlated neutron events only within a time gate of 300 ms from the ion event.

\begin{figure}[h]
\centering
\includegraphics[width=\columnwidth]{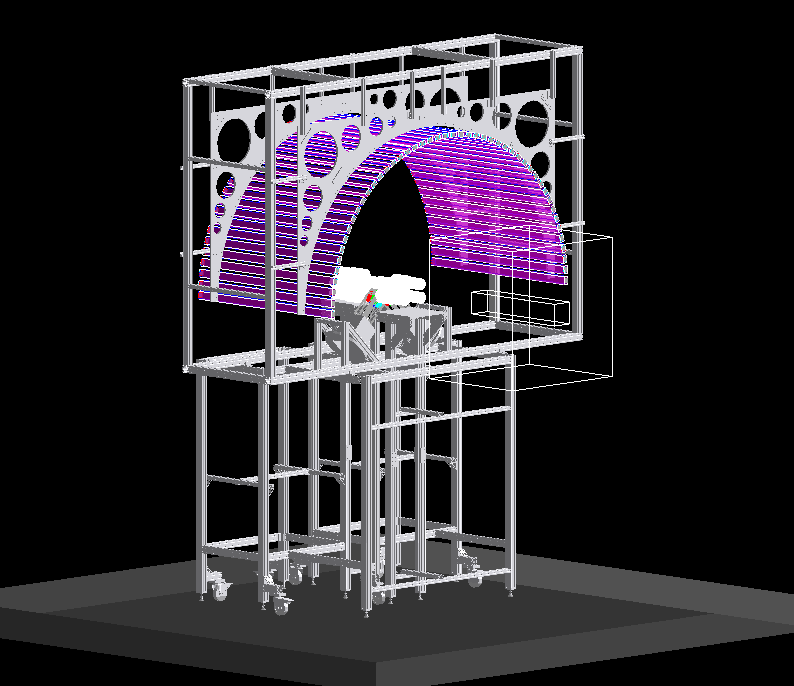}
\caption{Geometry of the experiment setup imported into the GEANT4. The setup consists of an 80/20 aluminum frame, two HPGe clovers, BRIKEN detector, ten \textit{LaBr\textsubscript{3}} detectors, concrete floor, and the implantation box.}
\label{fig:setup}
\end{figure}

For an ideal neutron time of flight measurement case, a Gaussian distribution is expected in the ToF spectrum corresponding to a given flight path of neutrons. However, in the presence of scattering materials from the point of origin to the point of detection of neutrons, the ToF distribution is expected to be asymmetric. The loss in energy due to scattering leads to tails in the ToF spectrum. To understand the scattering of neutrons in the experiment setup. A GEANT4 \cite{GEANT4, AGOSTINELLI2003250} simulation package was developed, which implements the interaction of neutrons and $\gamma$-rays with various types of materials that they may encounter. The simulations were performed to obtain the response of the VANDLE bars to mono-energetic neutrons with the addition of the implantation box containing the YSO detector. The timing resolution of the YSO was implemented in the simulation by using a Gaussian-shaped random generator with an FWHM of $\sim$ 650 ps. The geometry of other possible neutron scatterers at the F11 focal point at the RIKEN Nishina Center, which includes the BRIKEN detector and the concrete floor, was also imported into the simulation package. The geometry of the BRIKEN detector is approximated by an HDPE block having a cuboid cavity. The setup imported is shown in Figure \ref{fig:setup}, a snapshot of the GEANT4 graphical user interface (GUI). The simulation code also includes trace analysis methods for timing calculation of PMTs of the VANDLE array, which were gain-matched to the data to benchmark the QDC response of the VANDLE array.

\begin{figure}[h]
\centering
\includegraphics[width=\columnwidth]{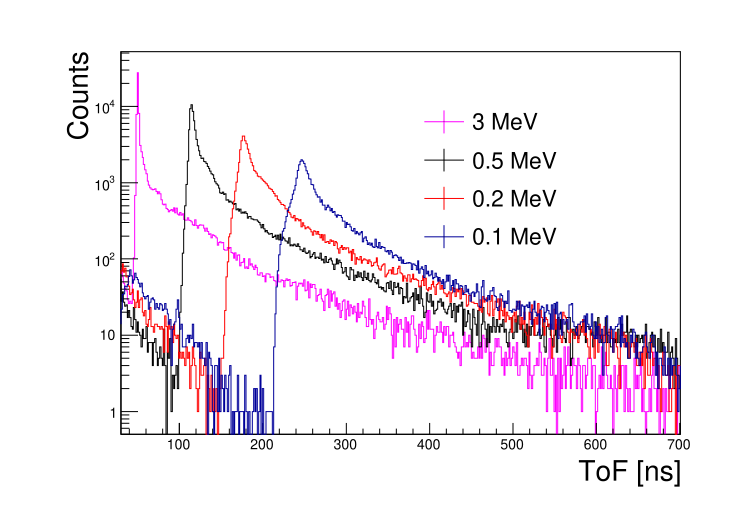}
\caption{VANDLE response to mono-energetic neutrons with various energy}
\label{fig:multi_energy_response}
\end{figure}

\begin{figure}[h]
\centering
\includegraphics[width=\columnwidth]{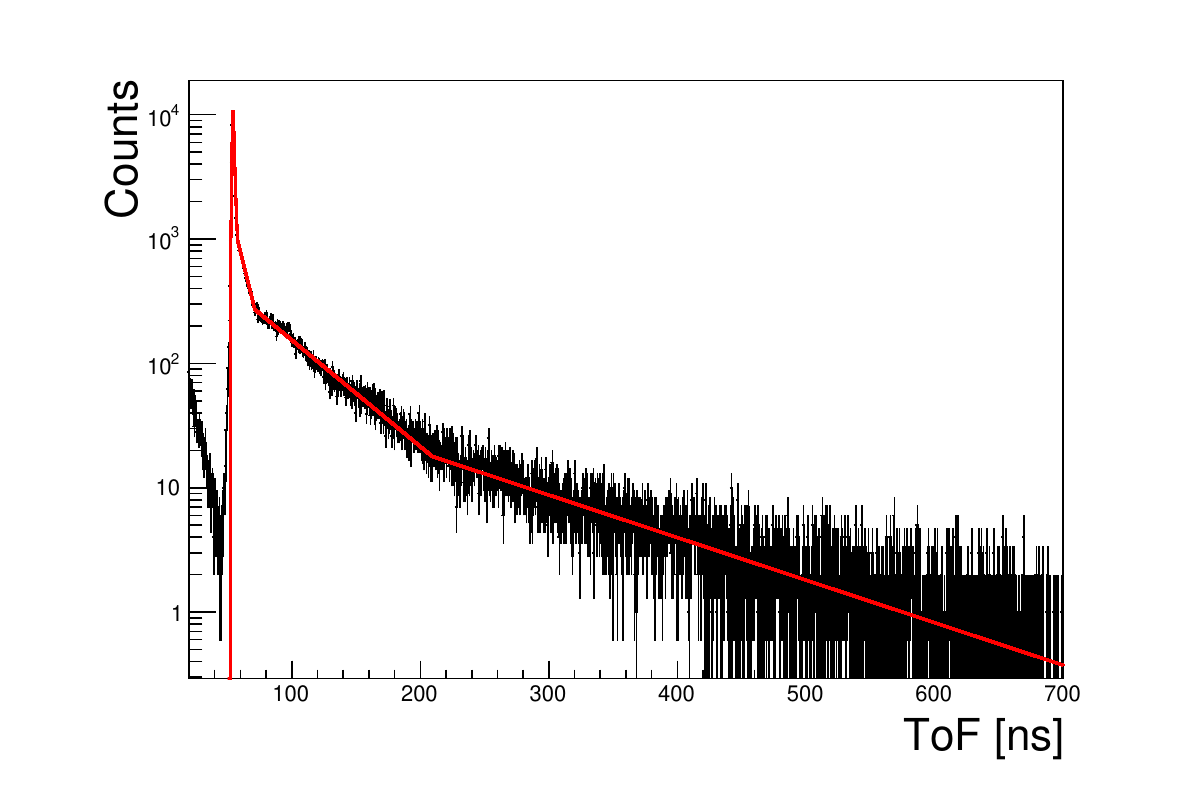}
\caption{VANDLE response to 2.5-MeV neutrons for the experiment at RIBF. The red line shows the response's $\chi$-square minimization fit with an asymmetric Lorentzian profile plus three exponential tails.}
\label{fig:2_5MeV_reponse}
\end{figure}

As mentioned earlier, the response of the VANDLE array to mono-energetic neutrons is affected by the amount of scattering experienced by neutrons on their way to the detection point in VANDLE from the point of origin is shown in Fig. \ref{fig:multi_energy_response}. Figure \ref{fig:multi_energy_response} shows the ToF distribution of various neutron energies, obtained using the GEANT4. The distributions show peak aroud ToF corresponding to energy for a flight path of 105-cm with tail like structures due to neutrons arriving late because of the scattering. The response function was fitted with an asymmetric Lorentzian profile plus three exponential tail components as shown in Figure \ref{fig:2_5MeV_reponse}, where the ToF distribution is fitted with a Breit-Wigner function plus three exponential tails. The parameters obtained by fitting the response function are used to extract neutron-energy peaks from a multi-neutron energy spectrum.

\section{Conclusion}
The YSO-based implantation detector is a compact and easy tool to measure half-life, and here we show that its use can be extended for ToF measurements for $\beta$-delayed neutrons when combined with fast-timing neutron detectors. The detector provides a high beta-detection efficiency and a fast timing response. The detector was used alongside VANDLE at the RIBF Riken Nishina Center, Japan, to measure beta-strength ($S_{\beta}$) to neutron-unbound states in the beta decay of r-process radioactive isotopes around doubly-magic \textsuperscript{78}Ni. As a demonstration, a QDC vs. TOF spectrum of $\beta$-delayed neutrons for \textsuperscript{83}Ga is presented, which has a signature typical of the VANDLE spectrum from previous measurements. A GEANT4 simulation toolkit has been prepared to study the response of the VANDLE array to capture the scattering of neutrons for de-convolving neutron TOF spectrum to calculate the intensity of neutron resonances. 


\section{Acknowledgement}
The experiment was carried out at the RI Beam Factory operated by RIKEN Nishina Center, RIKEN and CNS, University of Tokyo. This research was partly sponsored by the Office of Nuclear Physics, U.S. Department of Energy Ander Award No. DE-FG02-96ER40983 (UTK) and DE-AC05-00OR22725 (ORNL), and by the National Nuclear Security Administration under the Stewardship Science Academic Alliances program through DOE Award No. DE-NA0002132 and DE-NA0002934.

\bibliography{cas-refs}
\end{document}